\documentclass[10pt,conference]{IEEEtran}
\usepackage{graphicx} % Required for inserting images
\usepackage{caption}
\usepackage{subcaption}
\usepackage{quantikz}
\usepackage{cite}

\title{Quantum Reservoir Computing for Corrosion Prediction in Aerospace: A Hybrid Approach for Enhanced Material Degradation Forecasting}
\author{\IEEEauthorblockN{1\textsuperscript{st} Akshat Tandon}
\IEEEauthorblockA{ 
\textit{Airbus Central Research \& Technology}\\
Ottobrunn, Germany \\
akshat.tandon@airbus.com}
\and
\IEEEauthorblockN{2\textsuperscript{nd} James Brown}
\IEEEauthorblockA{ 
\textit{qBraid Co.}\\
Chicago, II, USA \\
james@qbraid.com}
\and
\IEEEauthorblockN{3\textsuperscript{rd} Kenny Heitritter}
\IEEEauthorblockA{ 
\textit{qBraid Co.}\\
Chicago, II, USA \\
kenny@qbraid.com}
\and
\IEEEauthorblockN{4\textsuperscript{th} Tarini Hardikar}
\IEEEauthorblockA{
\textit{qBraid Co.}\\
Chicago, IL, USA \\
tarini@qbraid.com}
\and
\IEEEauthorblockN{5\textsuperscript{th} Kanav Setia}
\IEEEauthorblockA{
\textit{qBraid Co.}\\
Chicago, IL, USA \\
kanavsetia@qbraid.com}
\and
\IEEEauthorblockN{6\textsuperscript{th} Rene Boettcher}
\IEEEauthorblockA{
\textit{Airbus Central Research \& Technology}\\
Ottobrunn, Germany \\
rene.boettcher@airbus.com}
\and
\IEEEauthorblockN{7\textsuperscript{th} Klaus Schertler}
\IEEEauthorblockA{
\textit{Airbus Central Research \& Technology}\\
Ottobrunn, Germany \\
klaus.schertler@airbus.com}
\and
\IEEEauthorblockN{8\textsuperscript{th} Jasper Simon Krauser}
\IEEEauthorblockA{
\textit{Airbus Central Research \& Technology}\\
Ottobrunn, Germany \\
jasper.krauser@airbus.com}
}

\begin{document}

\maketitle
\begin{abstract}
    The prediction of material degradation is an important problem to solve in many industries. Environmental conditions, such as humidity and temperature, are important drivers of degradation processes, with corrosion being one of the most prominent ones. Quantum machine learning is a promising research field but suffers from well known deficits such as barren plateaus and measurement overheads. To address this problem, recent research has examined quantum reservoir computing to address time-series prediction tasks. Although a promising idea, developing circuits that are expressive enough while respecting the limited depths available on current devices is challenging. In classical reservoir computing, the onion echo state network model (ESN) \cite{OESN} was introduced to increase the interpretability of the representation structure of the embeddings. This onion ESN model utilizes a concatenation of smaller reservoirs that describe different time scales by covering different regions of the eigenvalue spectrum. Here, we use the same idea in the realm of quantum reservoir computing by simultaneously evolving smaller quantum reservoirs to better capture all the relevant time-scales while keeping the circuit depth small. We do this by modifying the rotation angles which we show alters the eigenvalues of the quantum evolution, but also note that modifying the number of mid-circuit measurements accomplishes the same goals of changing the long-term or short-term memory. This onion QRC outperforms a simple model and a single classical reservoir for predicting the degradation of aluminum alloys in different environmental conditions. By combining the onion QRC with an additional classical reservoir layer, the prediction accuracy is further improved.
\end{abstract}

% some important comments 
% Change P1, P2, etc to something like env conditions in the graphs..
% graphs are too small and not readable either.
% Acknowledge the other people who helped in Data generation.

% "ACKNOWLDEGMENT" : SOmthing like: --->
% The authors gratefully acknowledge the contributions of Quentin Bignon from Airbus Central R&T, Nantes, France, as well as Christian Feiler, Bahram Vaghefinazari, Lisa Sahlmann and Mikhail Zheludkevich from Helmholtz Center Hereon, Geesthacht, Germany, for their efforts in generating the experimental data utilized in this study.

\section{Introduction}

Material degradation due to environmental exposure—such as corrosion, is a pervasive challenge across industries. Environmental conditions, including temperature, humidity, and electrolyte conductivity, are key drivers of corrosion processes. The onset and progression of corrosion are intricately linked to these macro-environmental factors, making it essential to establish a clear correlation between them and material degradation. This understanding is particularly critical in high-stakes sectors like aerospace, where maintaining material integrity and reliability directly impacts performance, and operational efficiency. Various AI methods have been used for the prediction of the evolution of corrosion prediction \cite{ruiz2024advanced, pourrahimi2023use,potnis2024, canonaco2022transfer, met13081459, s24113564}. In aerospace, such AI solutions can help improve material longevity, enable predictive maintenance, and support sustainability goals by reducing waste, optimizing manufacturing, and lowering costs.  Moreover, such AI solutions can enhance operational efficiency while contributing to the long-term reliability and environmental responsibility of the aircraft fleet.\cite{peng2024prediction}

Recently, many quantum variational machine learning algorithms have been explored for the prediction of time series data \cite{anand2024time, qmarl_akshat}. However, these algorithms often face challenges related to trainability, such as vanishing gradients and barren plateaus. To address these issues, quantum reservoir computing (QRC) has gained significant attention as an alternative approach. Unlike variational algorithms, QRC does not require gradient evaluation on quantum hardware, making it a promising candidate for such tasks\cite{beaulieu2024robust}. Quantum Reservoir Computing (QRC) adapts the principles of classical reservoir computing (CRC) to a quantum setting by using the complex dynamics of a quantum system, rather than a traditional dynamical network, to embed input data into a high-dimensional feature space. Because it bypasses gradient-based parameter updates on quantum hardware, QRC avoids the vanishing gradient problems that often plague variational quantum machine learning algorithms. The idea is that, through the native quantum interactions and measurements, one can generate a rich set of features for downstream tasks such as time-series forecasting, classification, and regression.

Although promising, realizing these benefits of QRC, in practice, requires sufficiently coherent quantum devices and thoughtful system design, as noise and limited qubit fidelity can limit the potential performance gains over classical approaches specially for time series prediction tasks. Also, the number of reservoir units, topology and spectral properties also crucial factors in producing an informative encoding of the input. In CRC, one approach to address some of these issues involves structuring reservoirs into segregated groups of units, where each group corresponds to an annular segment of the reservoir spectrum \cite{tortorella2024onion}. Drawing inspiration from this classical approach, we propose a hybrid Quantum Reservoir Computing framework that leverages multiple quantum reservoirs, with each reservoir representing an annular segment of the overall quantum reservoir spectrum, i.e having each reservoir units with different range of eigenvalues. This design aims to enhance encoding quality and improve the performance of QRC for time series prediction tasks by both improving the quality of reservoir and number of qubits used.

%[Different noise lead to different eignevlaues of reservoir ]
Previous research has shown\cite{Monzani2024} that different types of noise alters the eigenvalues of the circuit. Without noise or measurements, all of the eigenvalues are unit absolute value due to the inherent unitary nature of quantum gates. Here, we show that the eigenvalues of the evolution can be modified by utilizing either mid-circuit measurements or altering the hyper-parameters of the circuit. This is, practically, much more useful when developing quantum reservoirs as altering the noise inherent in the system is difficult. However, one may be able to take the noise parameters as a given for different hardware platforms and generate a robust reservoir by combining the reservoirs as described below.
%[Result]

\section{Material Problem}
\begin{figure}
    \centering
    \includegraphics[width=1\linewidth]{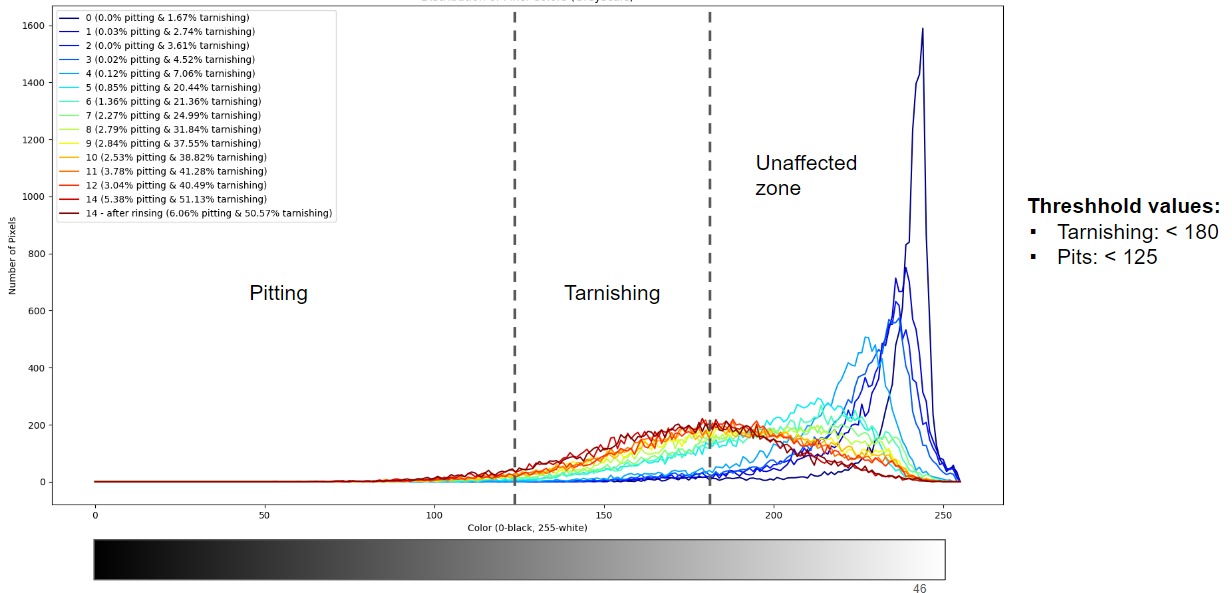}
    \caption{The definition of the tarnishing and pitting values that we are attempting to predict.Pixel color value distribution (gray-scale) of exposed samples and pixel value thresholds for pitting, tarnishing and uneffected zones that has been used to estimate the affected relative surface area values (Fig. \ref{fig:tarnishing_pitting_evolution}) }
    \label{fig:tp_def}
\end{figure}

\begin{figure}
     \centering
     \begin{subfigure}[b]{0.44\textwidth}
         \centering
         \includegraphics[width=\textwidth]{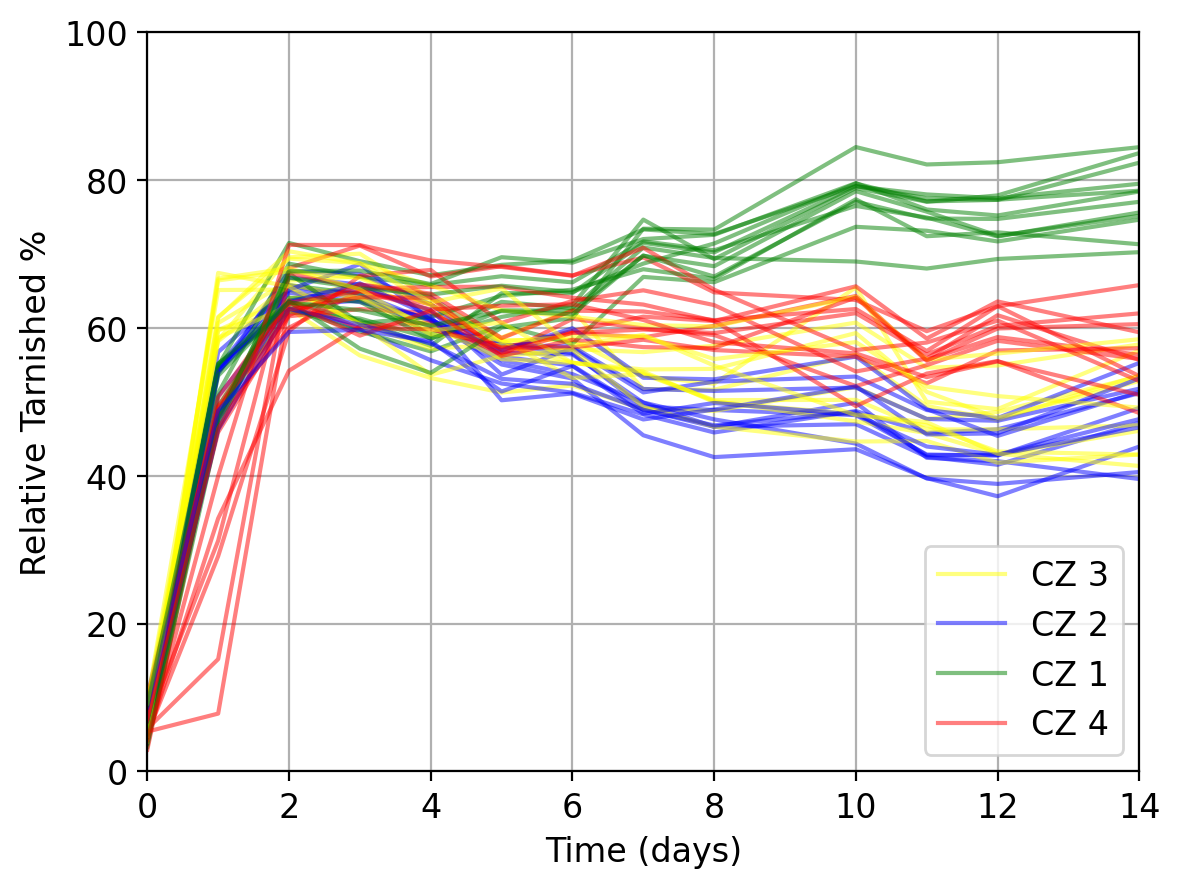}
         \caption{Tarnishing Evolution}
         \label{fig:tarnishing}
     \end{subfigure}
     \hfill
     \begin{subfigure}[b]{0.44\textwidth}
         \centering
         \includegraphics[width=\textwidth]{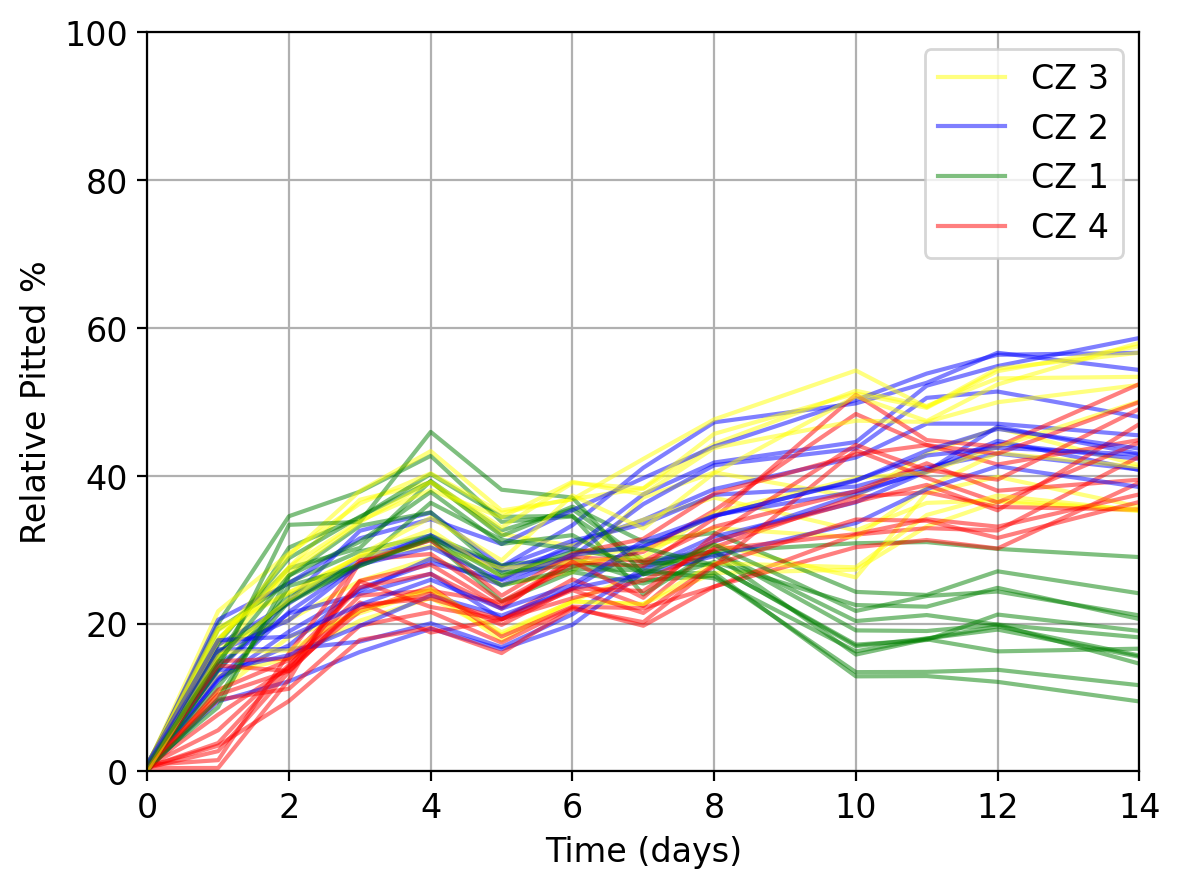}
         \caption{Pitting Evolution}
         \label{fig:pitting}
     \end{subfigure}
        \caption{The relative surface area affected by (above) tarnishing and (below) pitting for aluminum alloy samples exposed to cyclic environmental conditions within four different climate zones (CZ). }
        \label{fig:tarnishing_pitting_evolution}
\end{figure}

The dataset we use is aluminum placed in a corrosion container with four different climate conditions for 14 days. Every day, the sample is removed and an image is taken to examine the pitting and tarnishing that has evolved. The samples have been exposed in four climate zones (CZ) inside the climate chamber in order to achieve different impacts on the exposed samples. Each of the four CZ has a distinct pitting and tarnishing evolution pattern (Fig. 2).  These values are obtained by looking at the spectrum of the image and are defined as in Figure \ref{fig:tp_def} by taking the respective image under controlled lighting conditions to ensure comparability, followed by transforming the photograph into a gray-scale image and evaluating the relative pitted and tarnished area, respectively, based on the distribution of the pixel color values (Fig. 1). 
Each of the four environmental conditions induces a distinct evolution pattern, as can be seen in Figure \ref{fig:tarnishing_pitting_evolution}.
For each environmental condition zone (1 - 4), humidity and temperature along with conductance are measured via dedicated sensors that are located inside each environment condition zone. As can be seen in Figure \ref{fig:tarnishing_pitting_evolution}, the tarnishing and pitting measurements are highly correlated, and therefore we will restrict ourselves to predicting only pitting as the dynamics are more difficult to capture using simple techniques. Each environment contains 12 samples which we split into 10 for training and 2 for testing. The same reservoir structure is used for each environment but the linear output layer is individually trained. Although each dataset is small, part of the promise of QRC is that it is possible to obtain good results without access to large training sets\cite{beaulieu2024robust}.          

The prediction problem we will solve is, given the first 3 days of pitting measurements, to predict the next 10 days of corrosion only using the humidity and temperature values (i.e. values that can be obtained without touching the sample). This would allow companies to monitor the environmental conditions of a material remotely, potentially reduce the necessary testing duration drastically, leading to faster testing capabilities, and have a good estimate of when the material properties have degraded to the point that they should be inspected.

\section{Onion QRC}

Given an input signal $u_t$, an echo state network (a classical reservoir computing model) is defined by a matrix $W$. Starting with a state $x_0$ and matrix $M$. The echo state is evolved by repeating the operation $x_{t+1}=F(W x_{t}+u_t)$ where $F$ is some non-linear function. The desired output signal $s_t$ is then obtained at each time step as $s_t=W_{out}x_t+b$ where $W_{out}$ and $b$ define the output layer and are the only variables that are trained. The matrix $W$ is generally a random real matrix and is normalized such that the maximum absolute value of an eigenvalue is less than 1. This normalization is done to ensure the stability of the evolution.\cite{tortorella2024onion}

Trotorella et al\cite{tortorella2024onion} introduced a novel technique to decrease the expressibility and increase the explainability of the otherwise black-box reservoir. They noted that larger eigenvalues describe longer-time dynamics while smaller eigenvalues describe shorter-time dynamics. If one utilized a set of random matrices $W_i$ with its eigenvalue spectrum confined to a ring (i.e. $\epsilon_0 \leq \left|E_i\right|\leq \epsilon_1$ where $E_i$ are the eigenvalues and $\epsilon_0$ and $\epsilon_1$ are real numbers. One could fill the spectrum while utilizing smaller eigenvalues. The dynamics would then evolve according to a block diagonal matrix with entries $W_i$ describing different parts of the spectrum.
\begin{equation}\label{eq.OnionES}
    W=\left[\begin{array}{cccc}
         W_0&&&  \\
         &W_1&& \\
         &&\ddots \\
         &&&W_N
    \end{array}\right]
\end{equation}

To improve the expressibility and reduce the hardware requirements, we utilize this idea in the framework of quantum reservoir computing. In the current NISQ era, few operations can be performed before results are not useful due to noise. It has been shown previously\cite{Monzani2024} that different types of noise can produce different memory effects. We instead aim to tune the eigenvalue spectrum of the resulting evolution by changing hyper-parameters in the circuit. The two we will discuss are prefactors in rotations and mid-circuit measurements. 

As all (noiseless) quantum operations are unitary, it is necessary to introduce measurements to obtain eigenvalues that have an absolute value that is not one. These measurements can be performed at any point in the evolution of the quantum reservoir. Previously, this has been done through mid-circuit measurements\cite{Monzani2024} or a feed-forward model\cite{similarQRC}. To exemplify the effect of these measurements, we convert the measurements to the equivalent Kraus operator\cite{NielsenChuang} which for a single qubit is
\begin{equation}\label{eq.kraus}
    \rho^{\prime} = B_0 \rho B_0 + B_1 \rho B_1 
\end{equation}
where $B_0=diag(1,0)$  and $B_1 = diag(0,1)$ where $diag$ signifies that the matrix is diagonal with entries listed.

\begin{figure}
     \centering
     \begin{subfigure}[b]{0.22\textwidth}
         \centering
         \includegraphics[width=\textwidth]{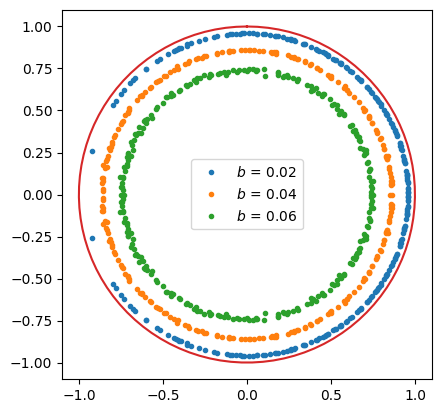}
         \caption{Increasing Rotation}
         \label{fig:moreangles}
     \end{subfigure}
     \hfill
     \begin{subfigure}[b]{0.22\textwidth}
         \centering
         \includegraphics[width=\textwidth]{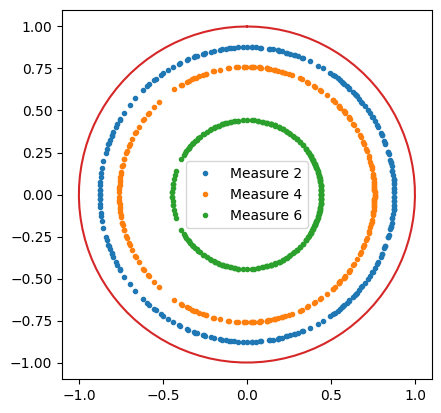}
         \caption{Increasing Measurements}
         \label{fig:moremeasurements}
     \end{subfigure}
        \caption{The eigenvalues of the evolution of a maximally mixed-state for the hardware efficient ansatz using differing rotation angles and mid-circuit measurements. a) It can be seen that increasing the rotation angle decreases the eigenvalues. One mid-circuit measurement on the first qubit is used.  b) More measurements decreases the eigenvalues for a constant rotation angle. By altering the rotation angles or the number of mid-circuit measurements, the onion-like layers of eigenvalues can clearly be viewed.}
        \label{fig:changing_eigenvalues}
\end{figure}

To most clearly show the layers of eigenvalues obtained by tuning hyperparameters to generate an onion QRC model, we use a hardware efficient ansatz\cite{HEA} with an additional layer of mid-circuit measurements. To obtain the eigenvalues of the evolution, we start with the maximally mixed state $\rho=I$ and evolve using the unitary gates and the non-unitary measurement Kraus operation of Eq. \ref{eq.kraus}. The rotation angles are fixed at a set of values with a prefactor that alters all rotations by a constant. According to Ref \citenum{similarQRC}, more rotation (larger prefactor) should decrease long-term memory and decrease the eigenvalues. As can be seen in figure \ref{fig:moreangles}, this is exactly what we see in this case. The layers of eigenvalues is clearly seen which is where the onion name comes from.

We also note that there is another nob that can be utilized to change the memory effects of the quantum reservoir. It can be seen in Figure \ref{fig:moremeasurements} that adding additional mid-circuit measurements also decreases the eigenvalues. When additional qubits are measured during the evolution of the system, the eigenvalues are decreased and therefore the memory is shifted more towards short-term memory.

Therefore, rotation prefactors and additional mid-circuit measurements can be utilized to tune the quantum reservoir to describe different timelines of memory. This could be combined with the knowledge of the noise channels, which also affect the memory\cite{Monzani2024}, to tune the quantum reservoir to specific Hardware backends. For our materials degradation use-case, we utilize the rotation angle effect to shift the eigenvalues of the evolution to better describe different time-frames of evolution.
 
\section{Models}
To begin with, we define a simple model that takes the average evolution of the 10 training samples in each environment and utilizes that to predict the evolution of the pitting values. 

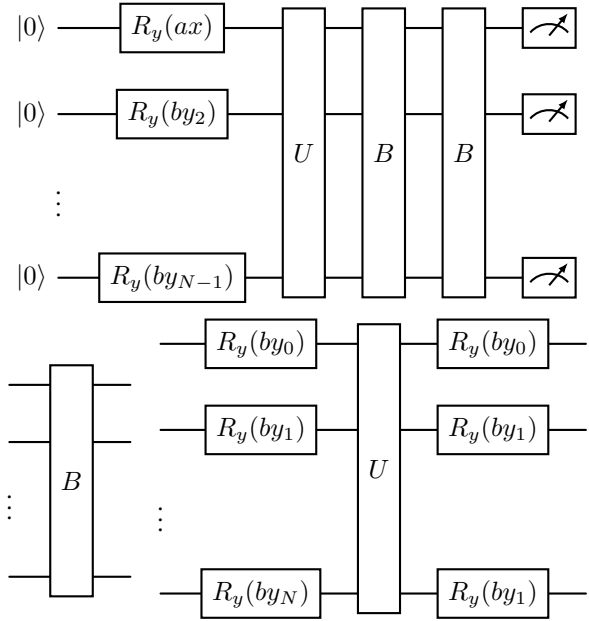
\begin{figure}
\begin{subfigure}[b]{0.5\textwidth}
\centering
\begin{quantikz}
\lstick{\ket{0}}&\gate{R_y(a x)}&\gate[4]{U}& \gate[4]{B}&\gate[4]{B}&\meter{} \\
\lstick{\ket{0}}&\gate{R_y(b y_2)}&&&&\meter{}\\
\vdots \\
\lstick{\ket{0}}&\gate{R_y(b y_{N-1})}&&&&\meter{}
\end{quantikz}
\end{subfigure}
\begin{subfigure}[b]{0.5\textwidth}
\centering
\begin{quantikz}[align equals at=2.5]
&\gate[4]{B}& \\
&& \\
\vdots \\
&& 
\end{quantikz}
\begin{quantikz}[align equals at=2.5]
&\gate{R_y(b y_0)} & \gate[4]{U}&\gate{R_y(b y_0)}& \\
&\gate{R_y(b y_1)}&&\gate{R_y(b y_1)}& \\
\vdots \\
&\gate{R_y(b y_N)}&&\gate{R_y(b y_1)}&
\end{quantikz}
\end{subfigure}
\caption{The feed-forward quantum reservoir that we utilize to predict corrosion. The measurements after a single step are converted into the $Z_i$ expectation value. $y_i = \mbox{arccos}(\langle Z_i\rangle)$ are then fed back into the circuit as the memory. The $U$ operation is a fully entangling layer of CNOTs as defined in Qiskit.}
\label{fig:qrc_circuit}
\end{figure}

The quantum gate-based model that we use is similar to that of Koboyashi et al\cite{similarQRC}. We utilize a feedback design where only the most recent predicted value is fed directly into the circuit while the memory of previous events is obtained by measuring the $Z$ expectation value of each qubit. There are multiple layers in the circuit described in Figure \ref{fig:qrc_circuit}. At each step the pitting percentage $x$ is fed into the circuit along with a prefactor of -0.30 times the normalized humidity and temperature. The previous $Z$ expectation values are fed into the circuit via $y_i = \mbox{arccos}(\langle Z_i\rangle)$. This was chosen as a circuit with only $R_y(y_i)$ on each qubit $i$ would have expectation value $\langle Z_i\rangle$.

By modifying the prefactor $b$ in the quantum reservoir, we can shift the eigenvalues to capture different timescales. When performing the prediction, we also obtain the expectation value of the pairs of $Z_{i}Z_{j}$ to be added to the readout layer. The onion QRC model is described by each individual circuit having its own feedback of $Z$ expectation values but the $x$ that is fed into each onion layer is the same. This $x$ value is predicted from the $Z_i$ and $Z_iZ_j$ combinations denoted $\langle Z_o\rangle $ for the $o^{th}$ layer by training the $C_{out}$ linear layer. This can be summarized as.
$x_t=C_{out} [\langle Z_1\rangle, \langle Z_2\rangle, ..., \langle Z_O\rangle ]$ where $\langle Z_o\rangle=C(a,b_o,x,\langle Z_{o}^{\prime}\rangle)$ where $\langle Z_{o}^{\prime}\rangle)$ is the expectation values from the previous layer and $C$ is the circuit of Figure \ref{fig:qrc_circuit}.

We also compare to a single classical reservoir computing (CRC) model defined by the equation
\begin{equation}
    X_{t+1} = \tanh \left(W_{in}\left[1,P_{t},h_{t},t_{t}\right]^{T}+Wx_{t}\right)
\end{equation}
where , $P_{t+1}=W_{out}\left[1,y_{t},h_{t},t_{t},x\right]^T$ is the pitting value at time $t+1$.

\section{Results}
To quantify the accuracy of the model, we utilize $R^2$ calculated using the $r2_{score}$ function of sklearn. For all models, the regularization parameter of the ridge regression is set to $1\times10^{-6}$. The simple model obtains an accuracy of 0.745. To examine the utility of the onion model, we optimize the hyperparameters of the circuit using 6 qubits. We obtain values of $a=-0.31$ and $b=0.1$. This is then set as fixed as we increase the number of qubits in the circuit. We now utilize the fact that the eigenvalues of the system can be shifted by increasing or decreasing the $b$ parameter as discussed above.

Before we continue, it should be noted that using one layer of the circuit design of Figure \ref{fig:qrc_circuit} improves the $R^2$ value with increasing number of qubits. As far as we know, this is the first time that a single set of hyper-parameters for a gate-based QRC model shows this property of increasing accuracy with increasing number of qubits. All previous work is special purpose few ($\leq 8$) qubits. This property has been seen with analog reservoirs\cite{scaleableQRC}.

\begin{figure}
    \centering
    \includegraphics[width=\linewidth]{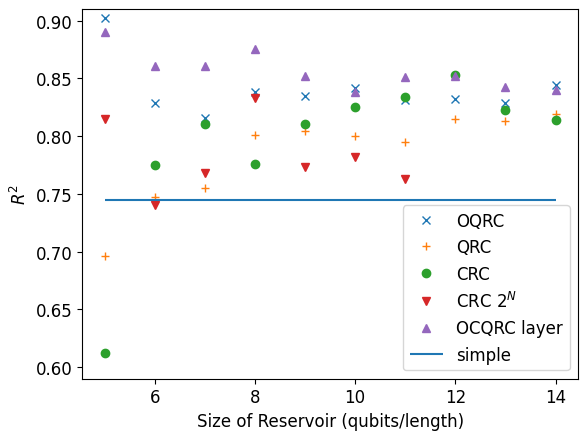}
    \caption{The utility of using onion-like ideas with QRC. As can be seen, additional layers (OQRC) tends to increase the $R^2$ value of the prediction.  The 3-layer QRC (OQRC) outperforms the 1-layer CRC for almost all qubit/length sizes. This is true whether one compares sizes with qubits equals length or qubits equals $2^N_q$ length where $N_q$ is the number of qubits. By making a hybrid classical onion quantum reservoir (OCQRC), even better results are obtained.}
    \label{fig:all_results}
\end{figure}

By adding more layers, we can see that the results tend to improve as is shown in Figure \ref{fig:all_results}. The 3 layer onion QRC (OQRC) also tends to do better than CRC for a range of classical reservoir sizes. This is also true for when the size of the classical reservoir is $2^{N_q}$ where $N_{q}$ is the number of qubits. We also performed calculations for 5-layer QRC and the results were improved but not significantly. Further eigenvalue tuning may increase the accuracy when adding more layers.

We also report a combined classical plus onion quantum reservoir (OCQRC) result in Figure \ref{fig:all_results}

 \section{Conclusion}
We have introduced a robust circuit design combined with the idea of onion QRC, and applied it to the prediction of material degradation.  In order to better utilize current hardware given that limits the realistic circuit depths one can use, we show how different timescales can be captured by modifying either rotation angles or the number of mid-circuit measurements. By simultaneously running fewer qubit quantum reservoirs and concatenating the measured expectation values, we can more accurately capture the necessary timescales while maintaining the simplicity of the circuit design.

We also show that by adding more layers, we can better predict the evolution of material degradation. In the future, it may be beneficial to combine classical and quantum reservoirs to better capture the dynamics. One could also combine different quantum modalities such as analog\cite{scaleableQRC} with the gate-based QRC utilized here. The methodology introduced here is highly flexible and should be useful in generating robust predictions of commercially important properties.

\section{Acknowledgments}
The authors gratefully acknowledge the contributions of Quentin Bignon from Airbus Central R\&T, Nantes, France, as well as Christian Feiler, Bahram Vaghefinazari, Lisa Sahlmann and Mikhail Zheludkevich from Helmholtz Center Hereon, Geesthacht, Germany, for their efforts in generating the experimental data utilized in this study and further providing  data analysis graph shown in Figure 1.
\bibliography{References}
\bibliographystyle{IEEEtran}
\end{document}